\documentclass[conference]{IEEEtran}

\def\review{1} 

\usepackage[T1]{fontenc}
\usepackage{blindtext}
\usepackage[utf8]{inputenc}
\usepackage[noadjust]{cite}
\usepackage[font=footnotesize]{caption}
\usepackage{dblfloatfix}
\usepackage{flafter}
\usepackage{graphicx}
\usepackage{psfrag}
\usepackage{subfig}
\usepackage{amsmath,amssymb,amsthm,mathrsfs,amsfonts,dsfont}
\usepackage{color}
\usepackage[algoruled]{algorithm2e}
\usepackage{fixltx2e}
\usepackage{multirow}
\usepackage{multicol}

\usepackage[hidelinks]{hyperref}
\usepackage{url}  

\usepackage[normalem]{ulem}
\usepackage{acronym}
\usepackage[table,xcdraw]{xcolor}
\usepackage{accents}
\usepackage{mwe}
\usepackage{hhline}
\usepackage{trfsigns}
\usepackage{siunitx}
\usepackage{lscape}
\usepackage{wasysym}
\usepackage{bm}
\usepackage{xfrac} 
\usepackage{url}

\if\review1
\usepackage{todonotes}
\else
\usepackage[disable]{todonotes}
\fi

\makeatletter
\newcommand*\bigcdot{\mathpalette\bigcdot@{.5}}
\newcommand*\bigcdot@[2]{\mathbin{\vcenter{\hbox{\scalebox{#2}{$\m@th#1\bullet$}}}}}
\makeatother

\usepackage{tikz}
\newcommand*\circled[1]{\tikz[baseline=(char.base)]{
            \node[shape=circle,draw,inner sep=1pt] (char) {#1};}}

\newacro{3GPP}{3rd Generation Partnership Project}
\newacro{5G}{fifth generation}
\newacro{5G NR}{Fifth Generation New Radio}
\newacro{6G}{sixth generation}
\newacro{A/D}{analog-to-digital}
\newacro{AAL}{array aperture line}
\newacro{ABE}{analog back-end}
\newacro{ADC}{analog-to-digital converter}
\newacro{AFE}{analog front-end}
\newacro{AGC}{automatic gain control}
\newacro{AGV}{automatic guided vehicle}
\newacro{AM-AM}{amplitude-to-amplitude modulation}
\newacro{AM-PM}{amplitude-to-phase modulation}
\newacro{AWGN}{additive white Gaussian noise}
\newacro{B5G}{beyond \ac{5G}}
\newacro{BB}{baseband}
\newacro{BER}{bit error ratio}
\newacro{BPSK}{binary phase-shift keying}
\newacro{BP}{band-pass}
\newacro{BS}{base station}
\newacro{CDM}{code-division multiplexing}
\newacro{CFO}{carrier frequency offset}
\newacro{CFR}{channel frequency response}
\newacro{CIR}{channel impulse response}
\newacro{CoMP}{coordinated multipoint}
\newacro{CP}{cyclic prefix}
\newacro{CPE}{common phase error}
\newacro{CPO}{carrier phase offset}
\newacro{CRLB}{Cram\'er-Rao lower bound}
\newacro{CS}{chirp sequence}
\newacro{CSI}{channel state information}
\newacro{CW}{continuous wave}
\newacro{CZT}{chirp Z-transform}
\newacro{D/A}{digital-to-analog}
\newacro{DAC}{digital-to-analog converter}
\newacro{DDC}{digital down-conversion}
\newacro{DDS}{direct digital synthesis}
\newacro{DFRC}{dual-function radar-communication or dual-functional radar-communication}
\newacro{DFnT}{discrete Fresnel transform}
\newacro{DFT}{discrete Fourier transform}
\newacro{DL}{downlink}
\newacro{DMRS}{demodulation reference signal}
\newacro{DoA}{direction-of-arrival}
\newacro{DoD}{direction-of-departure}
\newacro{DPD}{digital pre-distortion}
\newacro{DUC}{digital up-conversion}
\newacro{ETSI}{European Telecommunications Standards Institute}
\newacro{EVM}{error vector magnitude}
\newacro{FDE}{frequency-domain equalization}
\newacro{FDM}{frequency-division multiplexing}
\newacro{FO}{frequency offset}
\newacro{FR2}{Frequency Range 2}
\newacro{gNB}{gNodeB}
\newacro{HP}{high-pass}
\newacro{IBFD}{in-band full duplex}
\newacro{ICI}{intercarrier interference}
\newacro{IDFT}{inverse discrete Fourier transform}
\newacro{IDFnT}{inverse discrete Fresnel transform}
\newacro{IF}{intermediate frequency}
\newacro{IHE}{Institute of Radio Frequency Engineering and Electronics}
\newacro{I/Q}{in-phase/quadrature}
\newacro{IBO}{input back-off}
\newacro{IP1dB}{input-referred 1-dB compression point}
\newacro{ISAC}{integrated sensing and communication}
\newacro{ISI}{intersymbol interference}
\newacro{ISLR}{integrated-sidelobe level ratio}
\newacro{IoT}{Internet of Things}
\newacro{JCAS}{joint communication and sensing}
\newacro{KIT}{Karlsruhe Institute of Technology}
\newacro{KPI}{key performance indicator}
\newacro{LDPC}{low-density parity-check}
\newacro{LFSR}{linear-feedback shift register}
\newacro{LNA}{low-noise amplifier}
\newacro{LO}{local oscillator}
\newacro{LoS}{line-of-sight}
\newacro{LP}{low-pass}
\newacro{LS}{least squares}
\newacro{mmWave}{milimeter wave}
\newacro{MIMO}{multiple-input multiple-output}
\newacro{MLE}{maximum likelihood estimator}
\newacro{MLS}{maximum-length sequence}
\newacro{MRC}{maximal-ratio combining}
\newacro{MUSIC}{multiple signal classification}
\newacro{NAF}{normalized angular frequency}
\newacro{NB}{narrowband}
\newacro{NLoS}{non-line-of-sight}
\newacro{NR}{new radio}
\newacro{OCDM}{orthogonal chirp-division multiplexing}
\newacro{OFDM}{orthogonal frequency-division multiplexing}
\newacro{OOB}{out-of-band}
\newacro{OTA}{over-the-air}
\newacro{P/S}{parallel-to-serial}
\newacro{PA}{power amplifier}
\newacro{PACF}{periodic autocorrelation function}
\newacro{PAPR}{peak-to-average power ratio}
\newacro{PCCF}{periodic cross-correlation function}
\newacro{PLC}{powerline communication}
\newacro{PLL}{phase-locked loop}
\newacro{PMCW}{phase-modulated continuous wave}
\newacro{PMN}{perceptive mobile network}
\newacro{PN}{oscillator phase noise}
\newacro{PoC}{proof-of-concept}
\newacro{PPLR}{peak power loss ratio}
\newacro{PRBS}{pseudorandom binary sequence}
\newacro{PRS}{positioning reference signal}
\newacro{PSD}{power spectral density}
\newacro{PSF}{point spread function}
\newacro{PSLR}{peak-to-sidelobe level ratio}
\newacro{QPSK}{quadrature phase-shift keying}
\newacro{RadCom}{radar-communication}
\newacro{RCS}{radar cross section}
\newacro{RF}{radio-frequency}
\newacro{RFS}{random finite set}
\newacro{RIS}{reflective intelligent surface}
\newacro{RMS}{root mean square}
\newacro{RMSE}{root mean squared error}
\newacro{RX}{receiver}
\newacro{SC}[S\&C]{Schmidl \& Cox}
\newacro{SFO}{sampling frequency offset}
\newacro{SIC}{self-interference cancellation}
\newacro{SINR}{signal-to-interference-plus-noise ratio}
\newacro{SIR}{signal-to-interference ratio}
\newacro{SISO}{single-input single-output}
\newacro{SJ}{sampling jitter}
\newacro{SNR}{signal-to-noise ratio}
\newacro{SoC}{system-on-a-chip}
\newacro{SQNR}{signal-to-quantization-noise ratio}
\newacro{SSB}{synchronization signal block}
\newacro{STO}{symbol time offset}
\newacro{S/P}{serial-to-parallel}
\newacro{TDD}{time-division duplexing}
\newacro{TDE}{time-domain equalization}
\newacro{TDM}{time-division multiplexing}
\newacro{TDR}{time-domain reflectometry}
\newacro{TITO}{tilt inference of time offset}
\newacro{TO}{time offset}
\newacro{TR}{technical report}
\newacro{TS}{technical specification}
\newacro{TX}{transmitter}
\newacro{UE}{user equipment}
\newacro{UL}{uplink}
\newacro{ULA}{uniform linear array}
\newacro{V2V}{vehicle-to-vehicle}
\newacro{ZF}{zero forcing}
\newacro{ZP}{zero padding}

\makeatletter
\renewcommand*\env@cases[1][1.2]{%
	\let\@ifnextchar\new@ifnextchar
	\left\lbrace
	\def\arraystretch{#1}%
	\array{@{}l@{\quad}l@{}}%
}
\makeatother

\usepackage{fancyhdr}
\fancypagestyle{empty}{fancy}

\begin{document}

\title{Bistatic ISAC: Practical Challenges and Solutions}

\author{Lucas Giroto\IEEEauthorrefmark{1}, Marcus Henninger\IEEEauthorrefmark{1}, Alexander Felix\IEEEauthorrefmark{1}, Maximilian Bauhofer\IEEEauthorrefmark{2}, Taewon Jeong\IEEEauthorrefmark{3},\\ Umut Utku Erdem\IEEEauthorrefmark{3}, Stephan ten Brink\IEEEauthorrefmark{2}, Thomas Zwick\IEEEauthorrefmark{3}, Benjamin Nuss\IEEEauthorrefmark{4}, and Silvio Mandelli\IEEEauthorrefmark{1}\\
		\IEEEauthorblockA{\IEEEauthorrefmark{1}Nokia Bell Labs Stuttgart, Germany \\
        \IEEEauthorrefmark{2}Institute of Telecommunications, University of Stuttgart, Germany \\
        \IEEEauthorrefmark{3}Institute of Radio Frequency Engineering and Electronics, Karlsruhe Institute of Technology, Germany \\
        \IEEEauthorrefmark{4}Professorship of Microwave Sensors and Sensor Systems, Technical University of Munich, Germany \\
			E-mail: lucas.giroto@nokia-bell-labs.com
		}
	}

\maketitle

\begin{abstract}
	This article presents and discusses challenges and solutions for practical issues in bistatic integrated sensing and communication (ISAC) in 6G networks. Considering orthogonal frequency-division multiplexing as the adopted waveform, a discussion on system design aiming to achieve both a desired sensing key performance indicators and limit the impact of hardware impairments is presented. In addition, signal processing techniques to enable over-the-air synchronization and generation of periodograms with range, Doppler shift, and angular information are discussed. Simulation results are then presented for a cellular-based ISAC scenario considering system parameterization compliant to current 5G and, finally, a discussion on open challenges for future deployments is presented.
\end{abstract}


\IEEEpeerreviewmaketitle


\section{Introduction}\label{sec:introduction}

To exploit the inherently distributed nature of cellular networks, enabling bistatic \ac{ISAC} is an essential task for \ac{6G} networks. In the bistatic architecture, transmitter and receiver are widely separated. Besides greater sensing diversity and being a key enabler of multistatic \ac{ISAC} \cite{thomae2025}, this architecture avoids full-duplex challenges faced in the monostatic case \cite{mandelli2023}.

In spite of the aforementioned advantages, bistatic \ac{ISAC} introduces several challenges \cite{wymeersch2025}. These include required synchronization between transmitter and receiver to avoid sensing bias and overall performance degradation. Furthermore, the limited knowledge of the transmit data at the receiver for radar processing must also be accounted for. Finally, as in the monostatic case, hardware impairments must still be considered during system design in bistatic \ac{ISAC}. These include, e.g., antenna non-idealities, non-linear distortions, \ac{I/Q} imbalance, \ac{PN}, and further impairments in data converters.

\begin{figure}[!t]
	\centering
	
	\psfrag{AAAA}[c][c]{\scriptsize $R^\text{Tx---T}_p,\,f^\text{Tx---T}_{\mathrm{D},p}$}
	\psfrag{BBBB}[c][c]{\scriptsize $R^\text{T---Rx}_p,\,f^\text{T---Rx}_{\mathrm{D},p}$}
	\psfrag{CCCC}[c][c]{\scriptsize $R_0,\,f_{\mathrm{D},0}$}
	\includegraphics[width=\columnwidth]{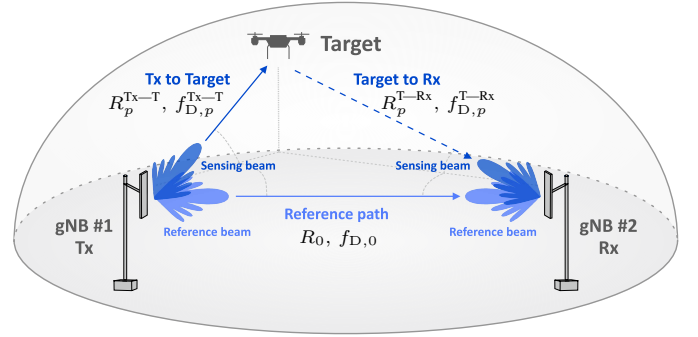}
	\captionsetup{justification=raggedright,labelsep=period,singlelinecheck=false}
	\caption{\ Bistatic ISAC system model. In this example, a static reference path labeled as $p=0$ and a path $p=1$ associated with a radar target are shown.}\label{fig:sysModel}
	\vspace{-0.5cm}
\end{figure}
	
	

In this article, these and further practical challenges in bistatic \ac{OFDM}-based \ac{ISAC} are described, and solutions in the contexts of system design and signal processing discussed.



\section{Notation and Link Budget}\label{sec:sys_model}

The considered bistatic \ac{ISAC} system is composed of two \acp{gNB} with static and known positions as depicted in Fig.~\ref{fig:sysModel}. It is assumed that \ac{gNB} \#1 acts as a transmitter and \ac{gNB} \#2 as a receiver. The \ac{OFDM} signal transmitted by \ac{gNB} \#1 occupies a bandwidth $B$ at a carrier frequency \mbox{$f_\mathrm{c}$} and is derived from the discrete-frequency domain frame \mbox{$\mathbf{X}\in\mathbb{C}^{N\times M}$} with \mbox{$N\in\mathbb{N}_{>0}$} subcarriers and \mbox{$M\in\mathbb{N}_{>0}$} \ac{OFDM} symbols. Before being received by \ac{gNB} \#2, this signal propagates through $P$ paths labeled as \mbox{$p\in\{0,1,\dots,P-1\}$}. The path $p=0$ is a reference path used for synchronization that also serves as a reference target for bistatic radar sensing, while the remaining $P-1$ paths are associated with radar point targets. While the reference is assumed to be a \ac{LoS} path in Fig.~\ref{fig:sysModel}, any dominant path, including \ac{NLoS} ones, with constant range, Doppler shift, and angle could be used for the same purpose. If the reference path is a non-dominant \ac{NLoS} one, the information that its parameters are constant can be used to help distinguishing it from paths associated with radar targets, therefore enabling the compensation for any eventual synchronization and target parameter estimation biases caused by a dominant radar target. 

The path $p$ is associated with delay \mbox{$\tau_p = R_p/c_0$}, where $c_0$ is the speed of light in vacuum and \mbox{$R_p=R^\text{Tx--T}_p+R^\text{T--Rx}_p$} is the bistatic range, which is equal to the sum of the range $R^\text{Tx--T}_p$ between \ac{gNB} \#1 and target and the range $R^\text{T--Rx}_p$ between target and \ac{gNB} \#2. In addition, a Doppler shift \mbox{$f_{\mathrm{D},p}=f^\text{Tx---T}_{\mathrm{D},p}+f^\text{T---Rx}_{\mathrm{D},p}$} is experienced. $f_{\mathrm{D},p}$ is the sum of the Doppler shifts $f^\text{Tx---T}_{\mathrm{D},p}$ and $f^\text{T---Rx}_{\mathrm{D},p}$ resulting from the relative movement of the target w.r.t. \ac{gNB} \#1 and \ac{gNB} \#2, respectively. Since the \acp{gNB} are static, it holds that \mbox{$f_{\mathrm{D},0}=\SI{0}{\hertz}$}. 
Denoting the complex baseband equivalent of the transmitted \ac{OFDM} signal by \ac{gNB} \#1 as \mbox{$x(t)\in\mathbb{C}$}, the baseband receive signal \mbox{$\tilde{y}(t)\in\mathbb{C}$} at \ac{gNB} \#2 is ideally expressed as
\begin{equation}\label{eq:yt}
	\tilde{y}(t) =~\alpha_0~x(t-\tau_0) + \sum_{p=1}^{P-1}\alpha_p~ x(t-\tau_p)~\e^{\im 2\pi f_{\mathrm{D},p}t}.
\end{equation}

\begin{table}[!t]
    \vspace{0.1in}
    \renewcommand{\arraystretch}{1.5}
    \arrayrulecolor[HTML]{708090}
    \setlength{\arrayrulewidth}{.1mm}
    \setlength{\tabcolsep}{4pt}
    		
    \centering
    \captionsetup{width=43pc,justification=centering,labelsep=newline}
    \caption{\textsc{Considered System Parameters}}
    \label{tab:fr2_params}
    \begin{tabular}{|c|c|}
        \hhline{|==|}
        \textbf{Carrier frequency ($f_c$)} & $\SI{27.4}{\giga\hertz}$ \\\hline
        \textbf{Frequency bandwidth ($B$)} & $\SI{190}{\mega\hertz}$ \\\hline
        \textbf{Subcarrier spacing ($\Delta f$)} & $\SI{120}{\kilo\hertz}$ \\\hline
        \textbf{Number of subcarriers ($N$)} & $1584$ \\\hline
        \textbf{Cyclic prefix length ($N_{\mathrm{CP}}$)} & $112$ \\\hline
        \textbf{OFDM symbols per frame ($M$)} & $1120$ \\\hline
        \textbf{Transmit power ($P_{\mathrm{Tx}}$)} & $\SI{36}{dBm}$ \\\hline
        \textbf{Tx/Rx antenna + array gain ($G_{\mathrm{Tx}},\,G_{\mathrm{Rx}}$)} & $\SI{33}{dBi}$ \\\hline
        \textbf{Noise figure ($NF$)} & $\SI{8}{dB}$ \\\hline
        \textbf{Number of ADC bits ($N_b$)} & $12$ \\\hline
        \textbf{Modulation alphabet ($\mathcal{M}$)} & QPSK \\\hhline{|==|}
    \end{tabular}
\end{table}

\begin{table}[!t]
    \renewcommand{\arraystretch}{1.5}
    \arrayrulecolor[HTML]{708090}
    \setlength{\arrayrulewidth}{.1mm}
    \setlength{\tabcolsep}{4pt}
        		
    \centering
    \captionsetup{width=43pc,justification=centering,labelsep=newline}
    \caption{\textsc{Sensing KPIs for Full-Frame Radar Signal Processing}}
    \label{tab:fr2_perf}
    \resizebox{\columnwidth}{!}{
    \begin{tabular}{|c|c|c|}
    	\hhline{|==|}
    	\textbf{Range resolution ($\Delta R$)} & $c_0/B =\SI{1.58}{\meter}$ \\ \hline
    	\textbf{Max. ua. range ($R_\mathrm{max,ua}$)} & $N~\Delta R = \SI{2.50}{\kilo\meter} $ \\ \hline
    	\textbf{Max. ISI-free range ($R_\mathrm{max,ISI}$)} & $N_\mathrm{CP}~\Delta R=\SI{176.65}{\meter}$ \\ \hline
    	\textbf{Doppler resolution ($\Delta f_\mathrm{D}$)} & $B/\left[\left(N+N_\mathrm{CP}\right)M\right]=\SI{100.07}{\hertz}$ \\ \hline
    	\textbf{Max. ua. Doppler ($f_\mathrm{D,max,ua}$)} & $\pm\,M\,\Delta f_\mathrm{D}/2=\pm\,\SI{56.04}{\kilo\hertz}$ \\ \hline
    	\textbf{Max. ICI-free Doppler ($f_\mathrm{D,max,ICI}$)} & $\pm\,\Delta f/10=\pm\,\SI{12.00}{\kilo\hertz}$ \\ \hline
        \textbf{Radar processing gain ($G_\mathrm{p}$)} & $N\,M=\SI{62.49}{dB}$ \\ \hhline{|==|}
	\end{tabular}}
    \vspace{-0.2cm}
\end{table}

In practice, \ac{STO} $\tau_\mathrm{STO}$ and \ac{CFO} $f_\mathrm{CFO}$ are experienced. In addition, several hardware impairments are experienced, and \ac{AWGN} impairs the receive signal. Representing the combined effect of these impairments as the operator \mbox{$\mathcal{F}_\mathrm{hw-imp}\{\cdot\}$} allows writing
\begin{equation}\label{eq:yt}
	y(t) = \mathcal{F}_\mathrm{hw-imp}\left\{\tilde{y}(t)\ast\delta(t-\tau_\mathrm{STO})\,\e^{\im 2\pi f_{\mathrm{CFO}}t}\right\}\,+\,n(t),
\end{equation}
where $y(t)$ is the complex baseband equivalent of the actual receive signal at \ac{gNB} \#2, $\ast$ is the convolution operator, and $n(t)$ is the \ac{AWGN}. 
After \ac{A/D} conversion with sampling frequency $f_\mathrm{s}$ and period \mbox{$T_\mathrm{s}=1/f_\mathrm{s}$} on $y(t)$, the sequence \mbox{$y[s]\in\mathbb{C}$} is produced. Due to \ac{SFO} \cite{giroto2024_tmtt} and \ac{SJ}, it holds that
\begin{equation}\label{eq:rx_sig_discrete_SFO}
	y[s] = y(t)\Big|_{t=sT_\mathrm{s}(1-\delta_\mathrm{SFO})+\tau_\mathrm{SJ}[s]}.
\end{equation}

After further processing on $y[s]$, a radar processing gain $G_\mathrm{p}$ is experienced, and the range-Doppler shift periodogram \mbox{$\mathbf{P}\in\mathbb{C}^{N\times M}$} produced. The latter can be expressed as
\begin{equation}
    \mathbf{P} = \tilde{\mathbf{P}} + \left(\mathbf{N}_\mathrm{AWGN} + \mathbf{P}_\mathrm{art} + \mathbf{P}_\mathrm{interf}\right).
\end{equation}
In this equation, \mbox{$\tilde{\mathbf{P}}\in\mathbb{C}^{N\times M}$} is the periodogram that is obtained under ideal synchronization and in the absence of hardware impairments. \mbox{$\mathbf{N}_\mathrm{AWGN}$} represents the \ac{AWGN} $n(t)$. The matrix \mbox{$\mathbf{P}_\mathrm{art}\in\mathbb{C}^{N\times M}$} contains the artifacts raised, e.g., by \ac{CPE} resulting from hardware impairments such as \ac{PN} \cite{giroto2024PN} or \ac{SJ} \cite{giroto2025_SJ}. Finally, the matrix \mbox{$\mathbf{P}_\mathrm{interf}\in\mathbb{C}^{N\times M}$} contains noise-like interference introduced by effects such as amplitude distortions or \ac{ICI} resulting from hardware impairments such as amplifier nonlinear distortion and \ac{I/Q} imbalance.

For the $p\mathrm{th}$ target in the periodogram $\mathbf{P}$, a link budget analysis leads to the maximum bistatic range parameter 
\begin{align}\label{eq:maxRange_bistatic}
    &\rho_p^\mathrm{max}=\sqrt{\left(R^\text{Tx---T}_p~R^\text{T---Rx}_p\right)^\mathrm{max}}\nonumber\\
    &= \sqrt[4]{\frac{P_{\mathrm{Tx}}~G_{\mathrm{Tx}}~G_{\mathrm{Rx}}~\sigma_p~\lambda^2~G_\mathrm{p}}{\left(4\pi\right)^3\,\left(k_\mathrm{B}\,B\,T_\mathrm{t}\,\mathrm{NF}+I_\mathrm{hw-imp}\right)\,\mathrm{SINR}^\mathrm{min}_\mathrm{per}}}.
\end{align}
In this equation, $P_\mathrm{Tx}$ is the transmit power, $G_\mathrm{Tx}$ and $G_\mathrm{Rx}$ are the transmit and receive antenna plus array gains, $\sigma_p$ is the \ac{RCS} of the the $p\mathrm{th}$ target, and \mbox{$\lambda=c_0/f_\mathrm{c}$} is the wavelength. Furthermore, \mbox{$k_\mathrm{B}B\,T_\mathrm{t}\,\mathrm{NF}$} is the \ac{AWGN} power, in which $k_\mathrm{B}$ is the Boltzmann constant, $T_\mathrm{t}$ is the standard room temperature in Kelvin, and $\mathrm{NF}$ is the receiver noise figure. $I_\mathrm{hw-imp}$ represents the combined power of all noise-like interference terms originating from hardware impairments, including quantization noise. Finally, $\mathrm{SINR}^\mathrm{min}_\mathrm{per}$ is the minimum required \ac{SINR} to detect target in the periodogram and perform unbiased parameters estimation, which is around $\SI{17}{dB}$ \cite{mandelli2023}. 



\section{System Design}\label{sec:sys_des}

For unbiased target parameter estimation from the periodogram $\mathbf{P}$, synchronization between \acp{gNB} \#1 and \#2 is required as later described in Section~\ref{sec:sync}. Additionally, system design must address the influence of \ac{OFDM} signal parameters on sensing \acp{KPI} and the effects of hardware impairments, as detailed in the following.

\subsection{Bistatic sensing KPIs}\label{subsec:sens_kpi}

The influence of \ac{OFDM} signal parameters on the maximum unambiguous values and resolutions for range and Doppler shift, the maximum \ac{ISI}-free range, the maximum \ac{ICI}-free Doppler shift, and the radar processing gain and target \ac{SINR} in the periodogram is discussed in \cite{brunner2024}. Considering the \ac{FR2} \ac{OFDM} parameterization based on numerology $\mu=3$ from the \ac{3GPP} \ac{TS} 38.211 for \ac{5G NR} \cite{3GPPTS38211}, the system parameters listed in Table~\ref{tab:fr2_params} are adopted. For the full-frame radar signal processing later explained in Section~\ref{sec:per_gen}, the resulting bistatic sensing \acp{KPI} listed in Table~\ref{tab:fr2_perf} are obtained.
%

\subsection{Effects of hardware impairments}\label{subsec:link_budget}

To define the maximum detectable range product $\rho_p^\mathrm{max}$ for a target with a given \ac{RCS} as described by \eqref{eq:maxRange_bistatic}, estimates of expected artifact and interference levels caused by hardware impairments and observed in the form of $\mathbf{P}_\mathrm{imp}$ and $\mathbf{P}_\mathrm{interf}$ are needed. The following items describes commonly experienced hardware impairments. It is worth highlighting that these are also relevant in monostatic architectures.

\subsubsection{Antenna/array non-idealities}

Antenna/array non-idealities include, e.g., antenna element coupling and beam squint. The first results in distortion of the ideal array patterns in both transmitting and receiving arrays, besides more severe spectral regrowth when combined with \ac{PA} nonlinearities at the transmitter. To reduce mutual coupling between antenna elements, countermeasures must be applied during array design. %
As for beam squint, it happens due to slightly different delays being experienced at the different \ac{BB} and \ac{RF} chains leading to the individual antennas. Its consequences are beamforming gain reduction, angular estimation bias, and inter-beam interference. Besides hardware-based countermeasures, signal processing approaches such as phase correction are possible.

\subsubsection{Amplifier non-linear distortion} This impairment is particularly relevant in \acp{PA} when their input signal covers a wide bandwidth and has high \ac{PAPR}, e.g., as in the case of \ac{OFDM} in \ac{5G NR}. If insufficient \ac{IBO} relative to the \ac{IP1dB} is not ensured, nonlinear \ac{AM-AM} and \ac{AM-PM} of the amplified signal are experienced. Consequently, in-band interference is experienced, besides spectral regrowth and resulting \ac{OOB} emissions specifically in the case of \acp{PA} at the transmitter.

\subsubsection{Mixer non-idealities} Besides similar non-linear distortion as in amplifiers, frequency-selective \ac{I/Q} imbalance can be experienced in quadrature mixers. This happens due to differences between the resulting \acp{CIR} of I and Q branches due to transfer functions of analog \ac{LP} filters as well as \acp{DAC} at the transmitter or \acp{ADC} at the receiver. Although \ac{I/Q} imbalance suppression measures can be taken during mixer design, further measures are required due to the high sensitivity of radar sensing. In \cite{schweizer2020}, a blind \ac{I/Q} imbalance estimation and correction approach for \ac{OFDM}-based radar systems is introduced, which can be directly applied to the considered bistatic \ac{OFDM}-based \ac{ISAC} system.
  
\subsubsection{Oscillator phase noise}\label{subsubsec:PN} \ac{PN} consists of phase variations experienced due to non-idealities in the \ac{LO} signal that is input to the mixers at both transmitter and receiver. In \ac{OFDM}-based systems, it results in \ac{CPE} and \ac{ICI}. The first impairs the \ac{DFT}-based Doppler shift estimation, leading to artifacts in the periodogram, target peak power loss, and estimation bias. 
As for \ac{ICI}, it results in loss of orthogonality between the subcarriers. Due to the random nature of transmit data, \ac{ICI} does not result in artifacts, but still leads to target peak power loss and increased interference level. %
To compensate these effects, \ac{PN} must be estimated at least in part. 
A feasible countermeasure consists of estimating and compensating the \ac{CPE} only \cite{giroto2024PN}, and leaving the \ac{PN}-induced \ac{ICI} to be naturally suppressed by the radar processing gain $G_\mathrm{p}$. In bistatic sensing, this can be done by estimating the phase of static reference paths \cite{griffiths2025}. Nevertheless, it must be considered that \ac{PN} has a \mbox{time-,} and therefore range-dependent behavior, which results in increased \ac{CPE} estimation error for targets associated with farther ranges than the reference path~\cite{giroto2024PN}. In \ac{MIMO} architectures, different phase offsets may be experienced in different receive chains. This is because different delays are experienced due to hardware tolerances leading to different \ac{LO} phases \cite{collmann2025}, besides slightly uncorrelated \ac{PN} realizations. Consequently, different \ac{CPE} is experienced at different receive chains, and \ac{DoA} estimation may be impaired.

\subsubsection{\ac{DAC}/\ac{ADC} quantization}

Quantization noise results from the finite resolution of data converters. Focusing on \acp{ADC}, as its associated quantization noise is more relevant for sensing link budget, the experienced \ac{SQNR} under perfect automatic gain control for a number \mbox{$N_\mathrm{b}\in\mathbb{N}_{>0}$} of \ac{ADC} bits and oversampling equal to \mbox{$f_\mathrm{s}/B$} is
\begin{equation}\label{eq:sqnr}
    \mathrm{SQNR~(dB)} = 6.02\,N_\mathrm{b} + 1.76 + 10\log_{10}(f_\mathrm{s}/B).
\end{equation}
Due to the randomness of transmit data in \ac{OFDM} signals, the radar processing gain $G_\mathrm{p}$ is also effective against quantization noise. Consequently, a factor \mbox{$10\log_{10}(G_\mathrm{p})$} is added to \eqref{eq:sqnr} to define the \ac{SQNR} in the periodogram.

\subsubsection{\ac{DAC}/\ac{ADC} clipping} The experienced clipping in \acp{DAC} and \acp{ADC} happens as soon as the input or output ranges of the converter are exceeded. This effect is known as hard clipping. In amplifiers, soft clipping happens in the transition between its linear and saturation regions. Only when in full saturation, amplifiers cause hard clipping of signals. Due to the similarity of the hard clipping in \acp{DAC} and \acp{ADC} and the full saturation in amplifiers, similar effects are expected during \ac{A/D} and \ac{D/A} conversion.
\subsubsection{\ac{DAC}/\ac{ADC} sampling jitter} \ac{SJ} manifests as random deviations from the ideal sampling points, and is primarily caused by \ac{PN} in the oscillators from which sampling clocks feeding \acp{DAC} and \acp{ADC} are derived. As discussed in \cite{giroto2025_SJ}, \ac{SJ} leads to \ac{ICI} and can also result in \ac{CPE} if \ac{BP} sampling is performed, as it impairs the \ac{DUC} and \ac{DDC} to and from the digital \ac{IF} frequency. Specifically at \acp{DAC}, \ac{SJ} may also lead to \ac{OOB} radiation 
Besides countermeasures such as \ac{CPE} compensation, jitter cleaning circuits can be used.%

\begin{figure*}[!t]
	\centering
	
	\psfrag{A}[c][c]{\footnotesize $y[s]$}
    \psfrag{B}[c][c]{\footnotesize $\mathbf{Y}$}
    \psfrag{C}[c][c]{\footnotesize $\hat{\mathbf{X}}$}
	\psfrag{D}[c][c]{\footnotesize $\mathbf{P}$}
	\psfrag{E}[c][c]{\scalebox{.65}{\footnotesize $\mathbf{G}$}}
	\psfrag{W}[c][c]{\scalebox{.65}{\footnotesize $\mathbf{W}^\mathrm{full}$}}
	
	\includegraphics[width=17cm]{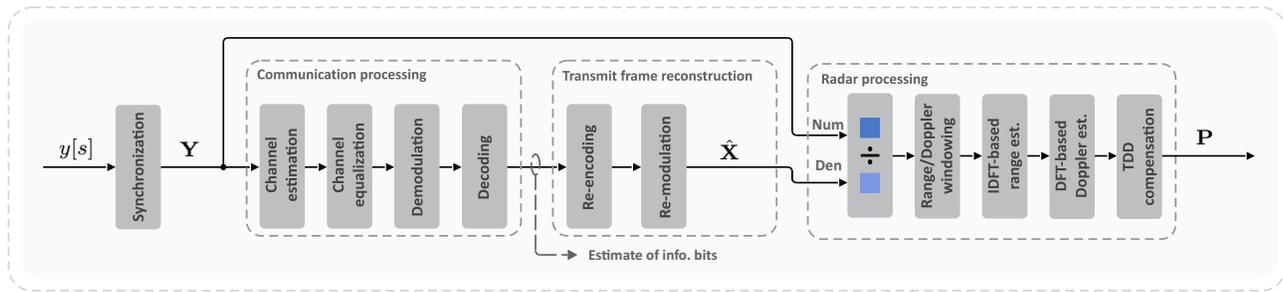}
	
	\captionsetup{justification=raggedright,labelsep=period,singlelinecheck=false}
	\caption{\ Bistatic OFDM-based ISAC receiver processing chain based on full-frame approach to obtain a range-Doppler shift periodogram.}\label{fig:bistaticRx_fullFrame}
\end{figure*}

\section{Signal Processing}\label{sec:sig_pro}

After system design, synchronization offsets between \acp{gNB} \#1 and \#2 must be accurately estimated and compensated, and the resulting signal processed for sensing accounting for limitations, e.g., due to cellular resource allocation. These aspects are covered in Sections~\ref{sec:sync} and \ref{sec:per_gen}. Additionally, angular estimation in \ac{MIMO} setups is discussed in Section~\ref{sec:ang_est}.

\subsection{Synchronization}\label{sec:sync}

To avoid periodogram distortion and enable unbiased target parameter estimation, \ac{STO}, \ac{CFO}, and \ac{SFO} have been identified as synchronization offsets to be estimated and compensated in \cite{brunner2024,giroto2024_tmtt}. It was shown that much finer offset estimation accuracy is required for bistatic sensing than for \ac{OFDM}-based communication systems. The algorithm developed over the course of these studies is constituted by the following steps.

\subsubsection{Coarse time and frequency synchronization} First, coarse \ac{STO} and \ac{CFO} estimates are obtained based on preambles (e.g., synchronization signal blocks in \ac{5G NR}), e.g., using cross-correlation based processing \cite{omri2019}. 
The \ac{CFO} is then corrected for a reduced sample range around the coarse \ac{OFDM} frame start point estimate obtained through the coarse \ac{STO} estimate.

\subsubsection{Sample-level time synchronization} A correlation of the resulting samples with one of the preamble \ac{OFDM} symbols is then performed to obtain an accurate \ac{STO} estimate, and therefore a sample-accurate \ac{OFDM} frame start sample estimate.

\subsubsection{Sampling frequency synchronization} 
Next, an \ac{OFDM} frame is formed and pilot subcarriers (e.g., carrying \ac{PRS} symbols) extracted. These pilots are then used to provide consecutive delay estimates for the reference path. Based on the delay drift over time, the \ac{TITO} algorithm \cite{giroto2024_tmtt} is used for robust and accurate \ac{SFO} estimation. Afterwards, the time-domain samples are resampled and a new \ac{OFDM} frame is formed.

\subsubsection{Time and frequency synchronization fine tuning} From the obtained frame, pilot subcarriers are again extracted and residual \ac{STO} and \ac{CFO} estimation is performed. At this stage, the residual delay and Doppler shift w.r.t. the reference path are estimated and corrected so that it is associated with zero range and Doppler shift. Consequently, biases in the relative range and Doppler shifts estimates for further targets are avoided.

\subsection{Periodogram generation}\label{sec:per_gen}

After synchronization, a discrete-frequency domain receive \ac{OFDM} frame \mbox{$\mathbf{Y}\in\mathbb{C}^{N\times M}$} is formed, on which multiple possible radar signal processing approaches can be adopted to generate range-Doppler shift periodograms. In \cite{brunner2024}, the pilot-based and full-frame approaches are discussed. The first consists of using only known pilots for radar signal processing, which comprises symbol division, \acp{IDFT} for range estimation and \acp{DFT} for Doppler shift estimation as in the monostatic case. The aforementioned known pilots can be, e.g., \ac{PRS}, and the fact that only a limited set of subcarriers is used results in limited sensing performance due to lower processing gain, maximum unambiguous range and maximum unambiguous Doppler shift. As for the latter approach, it consists in obtaining a full estimate \mbox{$\hat{\mathbf{X}}\in\mathbb{C}^{N\times M}$} of the transmit frame $\mathbf{X}$ via communication signal processing, allowing the use of the whole \ac{OFDM} frame for radar signal processing. Consequently, full processing gain, maximum unambiguous range and maximum unambiguous Doppler shift can be achieved, and superior sensing performance is attained. A requirement, however, is that little to no communication errors occur, which can be ensured via proper choice of modulation and coding schemes, e.g., as in \cite{henninger2026}. Otherwise, either an increased interference level or artifacts are observed in the periodogram if modulation symbol interleaving after encoding is performed or not, respectively \cite{brunner2024}. For both processing approaches, the radar performance parameters can be calculated as in \cite{brunner2024}.

In a cellular infrastructure-based \ac{ISAC} scenario, the \ac{TDD} between periodic \ac{DL} and \ac{UL} transmissions in \ac{3GPP}-compliant communication must be considered. If the \ac{DL} parts are patched together or the \ac{UL} parts are blanked, both a reduction of the processing gain $G_\mathrm{p}$ and artifacts along the Doppler shift direction of the periodogram will be observed. In \cite{henninger2025}, a method to reconstruct the point spread function of targets based on the \ac{TDD} pattern knowledge is proposed. This allows subtracting the contribution of individual point targets in the periodogram. Simulation and measurement results with a \ac{5G NR}-compliant \ac{ISAC} setup with commercial communication hardware in \ac{FR2} were presented, showing that the proposed method allows destinguishing between true targets and artifacts and cleaning the periodogram to enable detecting weaker targets.

To encapsulate the concepts detailed in Sections~\ref{sec:sync} and \ref{sec:per_gen}, Fig.~\ref{fig:bistaticRx_fullFrame} depicts the synchronization strategy and the periodogram generation procedure relying on full-frame processing. It is worth highlighting that the estimate of information bits shown in this figure can also be used for interference cancellation, e.g., to improve monostatic sensing or bistatic sensing involving another transmitting \ac{gNB}. More specifically, these bits can be used, alongside synchronization offset and channel estimates, to reconstruct the received discrete-time domain signal, which is then subtracted from the overall signal at the receiving \ac{gNB} of the bistatic pair, as described in \cite{jeong2025}.

\subsection{Angular domain sampling and reconstruction}\label{sec:ang_est}

%
To perform angular estimation under the constraint of beamformed transmission, one range-Doppler shift periodogram must be estimated per \ac{DoD}-\ac{DoA} pair. Minimal angular sampling, i.e., choice of a reduced set of \ac{DoD}-\ac{DoA} pairs, is especially necessary in setups with confined angular acquisition capabilities, such as those with hybrid or analog beamforming structures. As discussed in \cite{felix2025}, the changes of angular direction between consecutive pairs demonstrate that beamforming relies on distinct orthonormal bases for transmitter and receiver, which allows the task to be divided into two one-dimensional sampling problems for the azimuth-only case. For a single uniform linear array, the Fourier duality of the unitary antenna element domain and the normalized angular frequency domains align with \ac{DFT} beamforming for $\lambda/2$ spacing. As such, \ac{DFT}-based sampling and interpolation schemes can be applied on a per array basis to reconstruct the angular domain. With the reconstruction of a \ac{DoD}-\ac{DoA} region of interest, a four-dimensional periodogram (range, Doppler shift, \ac{DoD}, and \ac{DoA}) is obtained.
\subsection{Multi-target information fusion and tracking}\label{sec:track}
After range, angle and Doppler shift are estimated as discussed in Sections~\ref{sec:per_gen} and \ref{sec:ang_est}, they must be combined or fused. This can be achieved via tracking, which establishes a target state estimate over time, while mitigating false alarms and missed detections resulting, e.g., from clutter. In \cite{bauhofer2025_mot}, an unlabeled probability hypothesis density filter was applied to measured data in the range-Doppler shift domain. Sensing of up to six challenging target trajectories emulated with the \mbox{R\&S$\copyright$AREG800A} radar echo generator from Rohde \& Schwarz was performed with the same \ac{ISAC} setup mentioned in Section~\ref{sec:per_gen}. The adopted tracking approach was able to cover target birth and death and measurement-to-target association while considering resolution limitations, and a mean absolute ranging error smaller than $\qty{1.5}{\meter}$ was observed.

In outdoor deployments, only approximate placement and orientation of \acp{gNB} are known, e.g., due to survey errors and mounting tolerances. Since these parameters are needed to fuse range and angle into position estimates during tracking, inaccuracies eventually lead to target parameter estimation biases. While calibration based on reference targets is a potential solution, further investigation is still needed.

\section{Simulation Results}\label{sec:sim_res}

\begin{table}[!t]
    \vspace{0.1in}
    \centering
    \captionsetup{width=43pc,justification=centering,labelsep=newline}
    \caption{\textsc{Considered Impairments and Simulated Corresponding SIR}}
    \label{tab:impairments_effects}
    \begin{tabular}{|c|c|c|c|}
        \hhline{|====|}
        \textbf{Impairment} & \textbf{PPLR (dB)} & \textbf{Mean SIR (dB)} & \textbf{Min. SIR (dB)} \\\hhline{|====|}
        PA nonlinearity & $\SI{-0.09}{dB}$ & $\SI{100.31}{dB}$ & N/A \\\hline
        PN & $\SI{0}{dB}$ & $\SI{92.23}{dB}$ & $\SI{56.92}{dB}$ \\\hline
        ADC quantization & N/A & $\SI{129.72}{dB}$ & N/A \\\hline
        DAC/ADC SJ & $\SI{0}{dB}$ & $\SI{139.18}{dB}$ & $\SI{99.37}{dB}$ \\\hhline{|====|}
        \textbf{All combined} & \textbf{$\SI{-0.09}{dB}$} & \textbf{$\SI{91.60}{dB}$} & \textbf{$\SI{56.92}{dB}$} \\\hhline{|====|}
    \end{tabular}
\end{table}

To evaluate the effect of the discussed impairments on the sensing link budget for a bistatic \ac{OFDM}-based \ac{ISAC} system with the parameters from Table~\ref{tab:fr2_params}, \ac{PA} non-linear distortion, \ac{PN}, \ac{ADC} quantization, and \ac{SJ} were considered and simulated. The remaining impairments were either assumed to be avoided during system design or to be compensated via signal processing, which is feasible with the discussed approaches. For the \ac{PA}, a memoryless nonlinearity model based on a lookup table was used, and a $\SI{10}{dB}$ \ac{IBO} from the \ac{IP1dB} was considered. Regarding \ac{PN}, two independent realizations of the \ac{3GPP} model for \acp{gNB} at $\SI{30}{\giga\hertz}$ proposed in \cite{3GPPTR38803} were simulated for transmitter and receiver, respectively, as decribed in \cite{giroto2024PN}, yielding an integrated \ac{PN} level of $\SI{-32.09}{dBc}$. Due to the described reasons in Section~\ref{subsubsec:PN}, \ac{CPE} was not compensated. Furthermore, \mbox{$N_\mathrm{b}=12$} \ac{ADC} bits as in Table~\ref{tab:fr2_params}, besides \mbox{$f_\mathrm{s}=\SI{4}{\giga\hertz}$} and a digital \ac{IF} of $\SI{1}{\giga\hertz}$, which are, e.g., possible with the Zynq UltraScale+ RFSoC ZCU111, were assumed. The maximum signal level, which defines the quantization noise, was set as $\SI{20}{dB}$ higher than the average \ac{LoS} path power to account for its \ac{PAPR} and provide a margin. Finally, model parameters based on data from the data sheet for LMX2594 \ac{RF} synthesizer, which is used in ZCU111 \ac{SoC} platforms, were used to generate the \ac{PN} \acused{PSD}power spectral density shaping the experienced \ac{SJ} as in \cite{giroto2025_SJ}, while setting LMX2594's nominal \ac{SJ} \ac{RMS} level of $\SI{45}{\femto\second}$. Two independent realizations of the described process were used to simulate \ac{DAC} and \ac{ADC} \ac{SJ}.

\begin{figure}[!t]
    \vspace{0.1in}
		\centering

        \psfrag{AA}[c][c]{\tiny \circled{A}}
        \psfrag{BB}[c][c]{\tiny \circled{B}}
        \psfrag{CC}[c][c]{\tiny \circled{C}}
        
		\psfrag{0.15}[c][c]{\footnotesize $0.15$}
        \psfrag{1}[c][c]{\footnotesize $1$}
        \psfrag{2}[c][c]{\footnotesize $2$}
        \psfrag{3}[c][c]{\footnotesize $3$}
        \psfrag{4}[c][c]{\footnotesize $4$}
        \psfrag{5}[c][c]{\footnotesize $5$}
        \psfrag{6}[c][c]{\footnotesize $6$}

        \psfrag{-100}[c][c]{\footnotesize -$100$}
        \psfrag{-75}[c][c]{\footnotesize -$75$}
        \psfrag{-50}[c][c]{\footnotesize -$50$}
        \psfrag{-25}[c][c]{\footnotesize -$25$}
        \psfrag{0}[c][c]{\footnotesize $0$}
        \psfrag{25}[c][c]{\footnotesize $25$}
        \psfrag{50}[c][c]{\footnotesize $50$}	

        \psfrag{sqrt(R-Tx R-Rx) (km)}[c][c]{\footnotesize $\rho_p$\,(km)}
        \psfrag{Power in periodogram (dBm)}[c][c]{\footnotesize Power in periodogram\,(dBm)}
		\includegraphics[width=5.5cm]{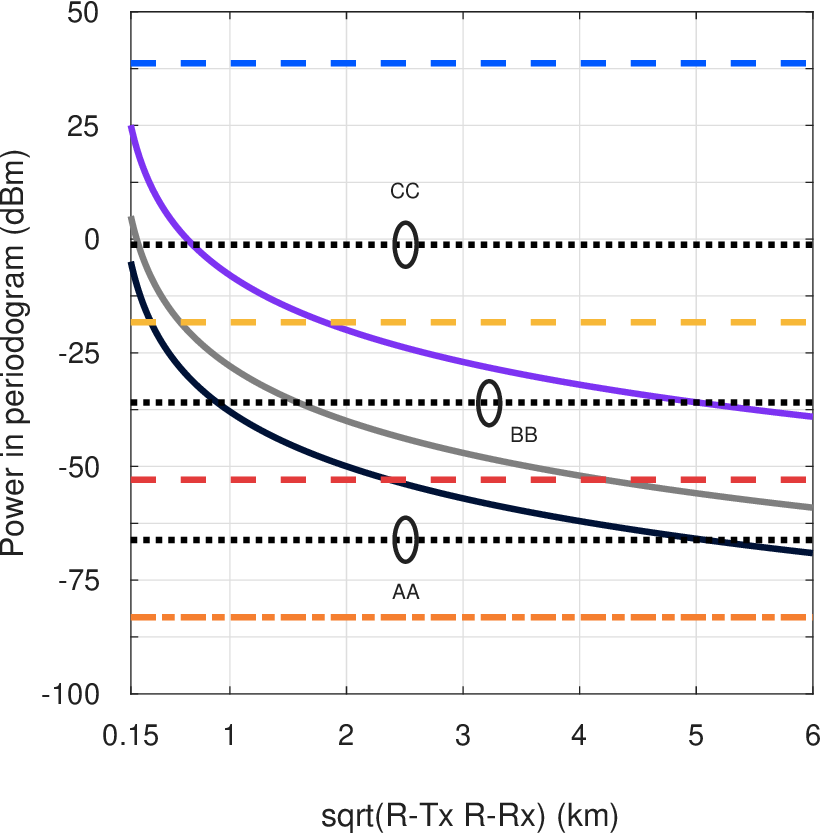}
		\captionsetup{justification=raggedright,labelsep=period,singlelinecheck=false}
		\caption{\ Power in periodogram accounting for antenna and processing gains as a function of the range parameter \mbox{$\rho_p=\sqrt{R^\mathrm{Tx-T}_p R^\mathrm{T-Rx}_p}$}. Results are shown for a drone ({\color[rgb]{0,0.07,0.21}\textbf{\rule[0.5ex]{1.25em}{1.25pt}}}), a pedestrian ({\color[rgb]{0.5,0.5,0.5}\textbf{\rule[0.5ex]{1.25em}{1.25pt}}}), and a car ({\color[rgb]{0.49,0.2,0.95}\textbf{\rule[0.5ex]{1.25em}{1.25pt}}}). In addition, the power of the \ac{LoS} path peak in the periodogram ({\color[rgb]{0,0.35,1}\textbf{\rule[0.5ex]{0.57em}{1.25pt}~\rule[0.5ex]{0.57em}{1.25pt}}}), its associated mean ({\color[rgb]{0.89,0.23,0.23}\textbf{\rule[0.5ex]{0.57em}{1.25pt}~\rule[0.5ex]{0.57em}{1.25pt}}}) and maximum ({\color[rgb]{0.97,0.72,0.22}\textbf{\rule[0.5ex]{0.57em}{1.25pt}~\rule[0.5ex]{0.57em}{1.25pt}}}) impairment-induced interference levels, and the \ac{AWGN} ({\color[rgb]{0.96,0.5,0.19}\textbf{\rule[0.5ex]{0.75em}{1.25pt}~\rule[0.5ex]{0.25em}{1.25pt}}}) level are shown. Finally, the power levels ({\color[rgb]{0,0,0}\textbf{\rule[0.5ex]{0.25em}{1.25pt} \rule[0.5ex]{0.25em}{1.25pt} \rule[0.5ex]{0.25em}{1.25pt}}}) yielding $\SI{17}{dB}$ \ac{SNR} against \ac{AWGN}, $\SI{17}{dB}$ \ac{SINR} against \ac{AWGN} and mean interference, and $\SI{17}{dB}$ \ac{SINR} against \ac{AWGN} and interference-induced artifacts are marked as \textcircled{\tiny A}, \textcircled{\tiny B}, and \textcircled{\tiny C}, respectively.}\label{fig:linkBudget}
\end{figure}
To obtain a sensing performance upperbound, \ac{DL}-only transmission was considered. \ac{PPLR} and periodogram \ac{SIR} for a scenario with a \ac{LoS} reference path and no targets were calculated assuming ideal synchronization. More specifically, a Chebyshev window with $\SI{100}{dB}$ sidelobe suppression was used for range and Doppler shift. Next, the \ac{PPLR} w.r.t. a impairment-free scenario with otherwise equal settings was calculated, and the \ac{SIR} was calculated as the ratio between the reference path peak and the region after a margin outside its main lobe. The obtained results are listed in Table~\ref{tab:impairments_effects}. For comparison, the procedures for calculating mean \ac{SIR} and minimum \ac{SIR} yield $\SI{153.19}{dB}$ and $\SI{100}{dB}$, respectively, for the impairment-free case due to the sidelobes of the reference path. 
Note that \ac{ADC} quantization noise is handled as a hardware impairment and its level is measured via \ac{SIR}, which is, however, equivalent to \ac{SQNR} as in \eqref{eq:sqnr}. Since it has a purely additive noise effect, \ac{PPLR} is not calculated for it. For \ac{PN} and \ac{SJ}, the minimum \ac{SIR} is shown, as these impairments cause both uniform interference and artifacts. To calculate of the combined effect of hardware impairments, the conservative assumption that these impairments are all uncorrelated was made. Consequently, the overall \ac{PPLR} is equal to the sum of the individual \acp{PPLR} in decibel scale, and the overall mean and minimum \ac{SIR} results from the sum of the respective mean and maximum interference levels in linear scale. 

Next, $\SI{21}{dBm}$ were allocated to the transmit beam associated with a \ac{LoS} reference path with a range of $\SI{300}{\meter}$, and the remaining $\SI{35.86}{dBm}$ to a sensing beam. Fig.~\ref{fig:linkBudget} shows the power levels in the periodogram for the reference path and three targets, namely a drone ($\SI{0.1}{\meter^2}$ \ac{RCS}), a pedestrian ($\SI{1}{\meter^2}$ \ac{RCS}), and a car ($\SI{100}{\meter^2}$ \ac{RCS}) \cite{mandelli2023}, all enhanced by the processing gain $G_\mathrm{p}$ minus the combined \ac{PPLR} of all impairments from Table~\ref{tab:impairments_effects}. While range and power for the reference path are fixed, the power levels of radar targets are shown as functions of the range parameter $\rho_p$ of the $p\mathrm{th}$ target. This parameter is defined as square root of the product of the range between transmitter and target and the range between target and receiver, i.e., \mbox{$\rho_p=\sqrt{R^\mathrm{Tx-T}_p R^\mathrm{T-Rx}_p}$}. In addition, levels for \ac{AWGN}, besides mean interference and artifacts due to hardware impairments, are also shown. Based on these results, the maximum $\rho_p$ for which \mbox{$\mathrm{SINR}^\mathrm{min}_\mathrm{per}\geq\SI{17}{dB}$} \cite{mandelli2023} was calculated with \eqref{eq:maxRange_bistatic}. The obtained results for $\rho_p^\mathrm{max}$ were $\SI{0.89}{\kilo\meter}$, $\SI{1.59}{\kilo\meter}$, and $\SI{5.01}{\kilo\meter}$ for drone, pedestrian, and car, respectively, with $I_\mathrm{hw-imp}$ defined by the mean interference level. In the impairment-free case, $\rho_p^\mathrm{max}$ is equal to $\SI{5.11}{\kilo\meter}$, $\SI{9.09}{\kilo\meter}$, and $\SI{28.75}{\kilo\meter}$, respectively, due to \ac{AWGN} only and disregarding the \ac{LoS} path sidelobes. If $I_\mathrm{hw-imp}$ is defined by the artifact level, the $\rho_p^\mathrm{max}$ values for drone, pedestrian, and car are reduced to $\SI{0.12}{\kilo\meter}$ (smaller than \mbox{$\rho_0=\SI{150}{\meter}$} for the reference path), $\SI{0.22}{\kilo\meter}$, and $\SI{0.68}{\kilo\meter}$, respectively. To avoid severe $\rho_p^\mathrm{max}$ limitation, the mean interference level can be set as the limiting factor, and artifacts can be handled, e.g., via \ac{CPE} compensation \cite{giroto2024PN} or clutter removal \cite{henninger2023}.

\section{Conclusion}\label{sec:conclusion}

This article provided an overview of practical challenges and solutions for \ac{6G} \ac{OFDM}-based bistatic \ac{ISAC}. In this context, the influence of \ac{OFDM} parameters on sensing \acp{KPI} and the impact of various hardware impairments were described, and key signal processing techniques discussed.

It was shown that, at first, accurate synchronization (\ac{STO}, \ac{CFO}, \ac{SFO}) needs to be achieved between \acp{gNB}. Then, periodogram generation with either pilots or knowledge of full transmitted frame was discussed, and practical issues due to \ac{TDD} patterns and angular domain sampling and reconstruction for \ac{DoD}-\ac{DoA} estimation considered. In addition, considerations on challenges such as mutual interference cancellation, multi-target information fusion and tracking, and the need for robust geometry calibration, were also made. Finally, simulation results showed that hardware impairments reduce sensing performance, limiting the maximum detectable range.

\section*{Acknowledgments}
The authors acknowledge the financial support by the Federal Ministry of Research, Technology and Space of Germany in the project KOMSENS-6G under grant numbers  16KISK112K, 16KISK113, and 16KISK010.

\bibliographystyle{IEEEtran}
\bibliography{./References/references}

\end{document}